\documentclass[twocolumn,showpacs,amsmath,amssymb,prd]{revtex4}

\usepackage{epsfig,epsf,graphics,psfrag}
\usepackage{times}
\usepackage{float}
\usepackage{color}
\usepackage{amsfonts,amssymb,stmaryrd,latexsym,amsmath}
\usepackage{multirow}
\usepackage{array} 
\usepackage{mathrsfs}
\usepackage{booktabs}
\usepackage{enumerate}
\usepackage{color}
\usepackage{subfigure}
\usepackage{hhline}
\usepackage[hyperindex=true]{hyperref}
\usepackage{epstopdf}

\usepackage{slashed}

\newcommand{\omeH}{\Omega_{cc}^{P,H}}
\newcommand{\omeL}{\Omega_{cc}^{P,L}}

\begin{document}

\title{New spectrum of negative-parity doubly charmed baryons:\\ Possibility of two quasistable states}

\author{Mao-Jun~Yan$^{1,2}$}
\author{Xiao-Hai~Liu$^{3,4}$}\email{xiaohai.liu@fz-juelich.de}
\author{Sergi~Gonz\`alez-Sol\'is$^1$}
\author{Feng-Kun~Guo$^{1,2}$}\email{fkguo@itp.ac.cn}
\author{Christoph~Hanhart$^3$}
\author{Ulf-G.~Mei{\ss}ner$^{4,3}$ }
\author{Bing-Song~Zou$^{1,2}$}

\affiliation{ 
$^1$CAS Key Laboratory of Theoretical Physics, Institute of Theoretical Physics,
Chinese Academy of Sciences, Beijing 100190, China\\
$^2$School of Physical Sciences, University of Chinese Academy of Sciences, Beijing 100049, China	\\
$^3$Institute for Advanced Simulation, Institut f\"ur Kernphysik and J\"ulich Center for Hadron Physics, Forschungszentrum J\"ulich,  D-52425 J\"ulich, Germany \\
$^4$ Helmholtz-Institut f\"ur Strahlen- und Kernphysik and Bethe Center for Theoretical Physics,
Universit\"at Bonn, D-53115 Bonn, Germany	
}

\date{\today}

\begin{abstract}

The discovery of $\Xi_{cc}^{++}$ by the LHCb Collaboration triggers
predictions of more doubly charmed baryons. By taking into account both the $P$-wave
excitations between the two charm quarks and the scattering of light pseudoscalar mesons off the ground state doubly charmed baryons, a set of
negative-parity spin-1/2 doubly charmed baryons are predicted already from a unitarized version of leading order
chiral perturbation theory. Moreover, employing heavy antiquark-diquark symmetry the relevant low-energy constants in the next-to-leading order are connected with those describing light pseudoscalar mesons scattering off charmed mesons, which have been well determined from lattice calculations and experimental data. Our calculations result
in a spectrum richer than that of heavy mesons. We find two very narrow $J^P=1/2^-$ $\Omega_{cc}^P$, which very likely decay into $\Omega_{cc}\pi^0$ breaking isospin symmetry. In the isospin-1/2 $\Xi_{cc}^P$ sector, three states are predicted to exist below 4.2~GeV with the lowest one being narrow and the other two rather broad. We suggest to search for the $\Xi_{cc}^{P}$ states in the $\Xi_{cc}^{++}\pi^-$ mode. Searching for them and their analogues are helpful to establish the hadron spectrum. 

\end{abstract}
 \maketitle

One of the most challenging problems in fundamental physics is to understand how
the strong interaction, formulated in terms of quantum chromodynamics (QCD),
organizes its spectrum observed as hadrons. The phenomenological constituent quark
 model achieved a great success in describing the majority of the hadron spectrum especially in the heavy quark sector
until 2003 when a few hadrons were discovered with unexpected properties. Since then many hadronic resonances beyond the conventional quark
model were discovered.

The new hadrons discovered in 2003 include the scalar
and axial-vector charm-strange mesons $D_{s0}^*(2317)$ and $D_{s1}(2460)$~\cite{Aubert:2003fg,Besson:2003cp}. Their masses are far below the
quark model predictions~\cite{Godfrey:1985xj}. The subsequent observations of
broad charm-nonstrange resonances $D_{0}^*(2400)$ and $D_{1}(2430)$~\cite{Abe:2003zm}
brought more puzzles. Thanks to the recent developments in lattice QCD
calculations of heavy-meson--light-meson
systems~\cite{Liu:2012zya,Mohler:2013rwa,Lang:2014yfa,Torres:2014vna,Lang:2015hza,Moir:2016srx,Bali:2017pdv}, to the precise experimental data of
$B^-\to D^+\pi^-\pi^-$~\cite{Aaij:2016fma}, and to the theoretical analysis of these lattice and
experimental data in the framework of unitarized chiral perturbation
theory~\cite{Liu:2012zya,Albaladejo:2016lbb,Du:2017zvv,Albaladejo:2018mhb,Guo:2018kno}, a consistent picture which can explain all the puzzles in these
positive-parity charmed mesons has emerged~\cite{Du:2017zvv}. In this picture, 
the  $D_{s0}^*(2317)$ and $D_{s1}(2460)$ are mainly $DK$ and $D^*K$ bound
states~\cite{Barnes:2003dj,vanBeveren:2003kd,Kolomeitsev:2003ac,Chen:2004dy,Guo:2006fu,Guo:2006rp}, respectively, and there are two nonstrange $0^+$ states and two $1^+$ states with isospin $I=1/2$ in the
ranges of the $D_0^*(2400)$ and $D_1(2430)$ masses, respectively.  According to the heavy quark flavor symmetry, all of these states have
their corresponding counterparts in the bottom meson spectrum. These low-lying positive-parity heavy mesons owe their existence to hadron-hadron interactions. This scenario needs to be checked against experimental and lattice results in other related processes, in order to reveal the proper paradigm of excited heavy hadrons.

The recent discovery of the doubly charmed baryon $\Xi_{cc}^{++}$ with a mass of $(3621.40\pm 0.78)$~MeV in $\Lambda_c^+K^-\pi^+\pi^+$ final states by the LHCb Collaboration~\cite{Aaij:2017ueg} opens new opportunities: First, this finding suggests the potential of discovering more low-lying doubly charmed baryons in the near future, and thus one needs to have a solid theoretical basis for the corresponding spectrum. Second, one would expect the positive-parity heavy mesons to have analogous counterparts as negative-parity doubly-heavy baryons, since the scattering of the pseudo-Nambu--Goldstone bosons (NGBs) ($\pi$, $K$ and $\eta$) off heavy sources is universal at leading order (LO). 
Moreover, employing an approximate symmetry of QCD even subleading terms can be fixed as detailed below.

For a doubly heavy baryon, the distance between the two heavy quarks $QQ$ may be estimated as $r_d\sim 1/(m_Q v_Q)$, with $v_Q$ the heavy quark velocity.
For an $S$-wave charm diquark one finds $m_cv_c\sim 800$ MeV~\cite{Hu:2005gf}.
On the other hand the distance of the light quark to the $QQ$ pair
is $r_q\sim 1/\Lambda_\text{QCD}$, with $\Lambda_\text{QCD}\sim 250$~MeV the scale of nonperturbative QCD. 
Thus one may expand in $r_d/r_q\sim 0.3$. To LO in this
expansion
the $S$-wave $QQ$ diquark appears as a point-like color antitriplet source, similar to a heavy antiquark, and this leads to an approximate heavy antiquark-diquark symmetry (HADS)~\cite{Savage:1990di}. Diquarks with higher partial waves are spatially much more extended, and such an approximation is not expected to work for them. This approximate symmetry allows one to predict doubly-heavy tetraquarks based on
input from heavy mesons as well as doubly and singly heavy baryons~\cite{Cohen:2006jg,Karliner:2017qjm,Eichten:2017ffp}
and, more relevant to our work, to relate
 doubly heavy baryons to singly heavy mesons~\cite{Savage:1990di,Georgi:1990ak,Carone:1990pv,Brambilla:2005yk,Fleming:2005pd, Hu:2005gf, Brodsky:2011zs,Ma:2015cfa,Mehen:2017nrh,Ma:2017nik}. Therefore, one can construct a chiral
effective field theory (EFT) describing the NGBs 
scattering off the ground state (positive-parity) doubly charmed baryons. The low-energy constants (LECs) in such a theory can be connected with those in the EFT describing NGBs scattering off ground state (negative-parity) anticharmed mesons. The latter has been extensively studied~\cite{Kolomeitsev:2003ac,Hofmann:2003je,Guo:2006fu,Guo:2009ct,Liu:2009uz,Altenbuchinger:2013vwa,Guo:2015dha,Liu:2012zya,Chen:2016spr,Yao:2015qia,Du:2017zvv,Guo:2018kno}.
In particular, the LECs in the next-to-leading-order (NLO) chiral Lagrangian
have been fixed by fitting to the lattice QCD results of several charmed-meson--light-meson $S$-wave scattering lengths~\cite{Liu:2012zya}, and the unitarized amplitudes using these inputs have been shown to be in a remarkable agreement with lattice QCD energy levels~\cite{Moir:2016srx} in the center-of-mass frame for the $S$-wave coupled channels $D\pi,D\eta$ and $D_s\bar K$~\cite{Albaladejo:2016lbb}, to be consistent with the lattice energy levels~\cite{Bali:2017pdv} for the $S$-wave $D^{(*)}K$~\cite{Albaladejo:2018mhb}, and to describe well the precise LHCb measurements~\cite{Aaij:2016fma} of the $D\pi$ angular moments for the decay $B^-\to D^+\pi^-\pi^-$~\cite{Du:2017zvv}. The predicted lowest positive-parity bottom-strange meson masses~\cite{Albaladejo:2016lbb} also agree nicely with the lattice QCD results~\cite{Lang:2015hza}. The existence of doubly charmed baryons analogous to the $D_{s0}^*(2317)$ has been proposed in Ref.~\cite{Guo:2011dd}, and was recently studied by considering potentials at LO ~\cite{Guo:2017vcf} or via light vector meson exchange~\cite{Dias:2018qhp}. 
In this Letter, in addition to using the NLO potentials, we notice that the $P$-wave excitations between the two heavy quarks have to be taken into account as dynamical degrees of freedom, leading to a distinct spectrum of novel states.

We consider the $S$-wave interactions between NGBs and the $J^P=1/2^+$ ground state doubly charmed baryons in the energy region around the corresponding thresholds. We are interested in the sectors with (strangeness, isospin) $(S,I)=(-1,0)$ and $(S,I)=(0,1/2)$, which have $\psi_{cc}\phi=\Xi_{cc}\bar K, \Omega_{cc}\eta$ and $\Xi_{cc}\pi, \Xi_{cc}\eta, \Omega_{cc} K$, respectively, as the relevant two-body coupled channels.
The coupled channel scattering amplitudes are collected in a $T$-matrix fulfilling unitarity, which can be written as~\cite{Oller:1997ng,Oller:1998hw,Oller:1998zr,Oller:2000fj,Oller:2000ma}
\begin{eqnarray}
\mathcal{T}(s)=\left[1-\mathcal{V}(s) G(s) \right]^{-1} \mathcal{V}(s),
\end{eqnarray}
where $s$ is center-of-mass energy squared. $G(s)$ is a diagonal matrix with the nonvanishing elements  $G_{ii}(s)=G(s,M_{\psi_{cc},i},M_{\phi,i})$ being the scalar one-loop function in the $i{\rm th}$ channel depending on the corresponding doubly charmed baryon and light meson masses $M_{\psi_{cc},i}$ and $M_{\phi,i}$. The loop function carries the unitary cut, and is calculated using a once-subtracted dispersion relation with the subtraction constant $a(\mu)$, where $\mu$ is an energy scale,~\cite{Oller:1998zr} serves as a regulator of the ultraviolet divergence.
The matrix $\mathcal{V}(s)$ stands for the $S$-wave projection of the potentials. It is split into two parts $\mathcal{V}(s) = \mathcal{V}_c(s)+\mathcal{V}_s(s)$. $\mathcal{V}_c(s)$ represents  the contact terms derived from the chiral Lagrangian up to NLO taking a similar form as that for the charmed mesons~\cite{Guo:2008gp,Guo:2009ct,Liu:2012zya} with the charmed meson fields replaced by those of the doubly charmed baryons. The HADS relates the involved LECs ($c_{0,1,...,5}$) to those in the charmed meson Lagrangian ($h_{0,1,...,5}$), as can be easily worked out
with the superfield formalism of Refs.~\cite{Hu:2005gf,Mehen:2017nrh}:
\begin{eqnarray}\label{LECsRelation}
c_i=\frac{h_i}{2\bar{M}_D},\ i=0,\ 1,\ 24,\ 35,
\end{eqnarray}
where $c_{24}=c_2+c_4 \bar{M}^2_{\psi_{cc}}$ and $c_{35}=c_3+c_5 \bar{M}^2_{\psi_{cc}}$. Here, $\bar M_D$ and $\bar M_{\psi_{cc}}$ are the averaged masses of the ground state charmed mesons and doubly charmed baryons, respectively.
For recent studies of doubly charmed baryons in chiral perturbation theory, we refer to Refs.~\cite{Sun:2014aya,Sun:2016wzh,Yao:2018ifh}. Furthermore, $\mathcal{V}_s(s)$ contains $s$-channel doubly charmed-baryon exchange potentials as discussed below.

The lowest excitations of doubly charmed baryons are due to the $P$-wave excitation inside the $cc$ diquark. Since the potential inside the color antitriplet $cc$ diquark is believed to be half of that between the $c$ and $\bar c$ in a charmonium, one expects that the $P$-wave excitation energy is roughly half of that for charmonia~\cite{Mehen:2017nrh}, {\it i.e.}, ${M}_{\psi_{cc}^P}-M_{\psi_{cc}}\simeq (M_{h_c}-M_{J/\psi})/2 = 214$~{MeV}, where $\psi_{cc}^P$ denotes the doubly charmed baryons with a $P$-wave diquark excitation. This value is similar to that calculated in quark models, see, e.g., Refs.~\cite{Ebert:2002ig,Kiselev:2001fw,Gershtein:2000nx}. With the excitation energy being of $\mathcal{O}(M_\pi)$, the $\psi_{cc}^P$ baryons have to be included explicitly as dynamical degrees of freedom. Therefore,  for a proper description of the low-energy $\psi_{cc}\phi$ interactions, we need the $S$-wave coupling~\cite{Hu:2005gf,Mehen:2017nrh}
\begin{eqnarray}
\label{s-channel-Lagrangian}
\mathcal{L}_P=\lambda \bar{\psi}_{cc}^P \gamma^\mu  u_\mu^{} \psi_{cc}^{}+h.c.,
\end{eqnarray}
where $\psi_{cc}^P=(\Xi_{cc}^{P++},\ \Xi_{cc}^{P+},\ \Omega_{cc}^{P+})^T$ represents the doubly charmed baryons with a $P$-wave $cc$ diquark, and $u_\mu = -\sqrt{2}\partial_\mu\phi/F_0+\mathcal{O}(\phi^3)$ is the axial current. Here, $F_0$ denotes the pion decay constant in the chiral limit, and $\phi=\sum_{i=1}^{8}\lambda_i\phi^i/\sqrt{2}$, with $\lambda_i$ the Gell-Mann matrices, collects the SU(3) NGB octet. Fermi statistics fixes the total spin of the $cc$ diquark in the ground state $\psi_{cc}$ and in the $\psi_{cc}^P$ to be $1$ and 0, respectively. Thus, the transition $\psi_{cc}^P \to \psi_{cc}\phi$ needs a flip of the charm quark spin, breaking heavy quark spin symmetry, and the dimensionless coupling constant $\lambda$ should be $\lambda=\mathcal{O}(\Lambda_\text{QCD}/m_{c})\ll 1$. The  tree-level amplitude for $\psi_{cc}^i(p_1)\phi^i(p_2)\to \psi_{cc}^f(p_3)\phi^f(p_4)$ from exchanging a $\psi_{cc}^P$ reads
\begin{equation}
V_s = \frac{2\lambda^2}{F_0^2}\mathcal{C}^{(s)} \bar{u}^f(p_3, \sigma^f) \slashed{p}_4 \frac{1}{\slashed{P} -\mathring{M}_{\psi_{cc}^P}} \slashed{p}_2 u^i(p_1, \sigma^i) , ~~~
\end{equation}
where $\sigma^i$ ($\sigma^f$) indicates the polarization of the initial (final) state baryon, $P=p_1+p_2=p_3+p_4$, and the coupled channel coefficients $\mathcal{C}^{(s)}$ are given in matrix form as
\begin{equation}
 \begin{pmatrix}
 2 & -\frac{2}{\sqrt{3}} \\
 -\frac{2}{\sqrt{3}} & \frac{2}{3}
 \end{pmatrix},\quad
 \mbox{and}\ \
 \begin{pmatrix}
 \frac{2}{3} & \frac12 & \frac{\sqrt{6}}{2} \\
 \frac12 & \frac16 & \frac{1}{\sqrt{6}}\\
 \frac{\sqrt{6}}{2} & \frac{1}{\sqrt{6}} & 1
 \end{pmatrix}
\end{equation}
for $(S,I)=(-1,0)$ and $(S,I)=(0,1/2)$, respectively.
The $S$-wave projection of $V_s$ gives the elements of the matrix $\mathcal{V}_s(s)$. It is worthwhile to notice that, analogous to the charmed meson case~\cite{Cleven:2010aw,Du:2017ttu}, the $u$-channel exchange of doubly charmed baryons gives a negligible contribution to the $S$-wave $\psi_{cc}\phi$ scattering, as checked in Ref.~\cite{Guo:2017vcf}.

The values of LECs are fixed from Eq.~\eqref{LECsRelation}.
The values of the $h_i$ have already been fixed from fitting to the lattice results for several charmed-meson--NGB scattering lengths at a few pion masses~\cite{Liu:2012zya}, which lead to the prediction of $2317^{+18}_{-28}$~MeV for the mass of the $D_{s0}^*(2317)$. 
Using the matching prescription in Refs.~\cite{Guo:2006fu,Yao:2018tqn}, the subtraction constant $a(\mu)$ in the charmed meson sector~\cite{Liu:2012zya} is translated to the doubly charmed sector as $a^{\psi_{cc}\phi}(1\ \mbox{GeV})=-2.79^{+0.04}_{-0.05}$.

As input for the hadron masses we take the isospin averaged values for all the mesons involved and use 3621.4~MeV~\cite{Aaij:2017ueg} for the $\Xi_{cc}$.
For the ground state ${\Omega_{cc}}$ we use a mass of $3725$~MeV fixed by requiring $M_{\Omega_{cc}^+}-M_{\Xi_{cc}^+}=M_{D_s^+}-M_{D^+}$ from HADS~\cite{Brodsky:2011zs}.
The quark model prediction from Ref.~\cite{Ebert:2002ig}, which correctly predicted the $\Xi_{cc}$ mass, is used as the bare mass of $\Xi_{cc}^P$, {i.e.}, $\mathring{M}_{\Xi_{cc}^P}=3838$~MeV, corresponding to the $P$-wave diquark excitation energy being 217~MeV. And we use $\mathring{M}_{\Omega_{cc}^{P}}\simeq M_{\Omega_{cc}}+217$~MeV$\simeq3942$~MeV. The symbol $\mathring{M}$ is used to emphasize that these values are the bare masses for the $1/2^-$ states without the $\psi_{cc}\phi$ dressing, to be distinguished from the pole masses from the coupled channel dynamics in the following. 
The only free parameter is the coupling $\lambda$ in the $s$-channel potential. 

The masses and widths of the low-lying $1/2^-$ doubly charmed baryons can be obtained by searching for poles of the coupled channel $T$-matrix with the corresponding quantum numbers. Depending on the channels and parameters, there can be real bound state poles in the first Riemann sheet of the complex energy plane, and/or poles in the second Riemann sheet (corresponding to a virtual state if the pole is real and below threshold, and a resonance if the pole is complex). The position of a real pole gives the mass of a physical state, and for a resonance, the pole is denoted as $M-i\,\Gamma/2$ with $M$ the mass and $\Gamma$  the width.

\begin{figure}[tb]
\centering
  \includegraphics[width=0.95\linewidth]{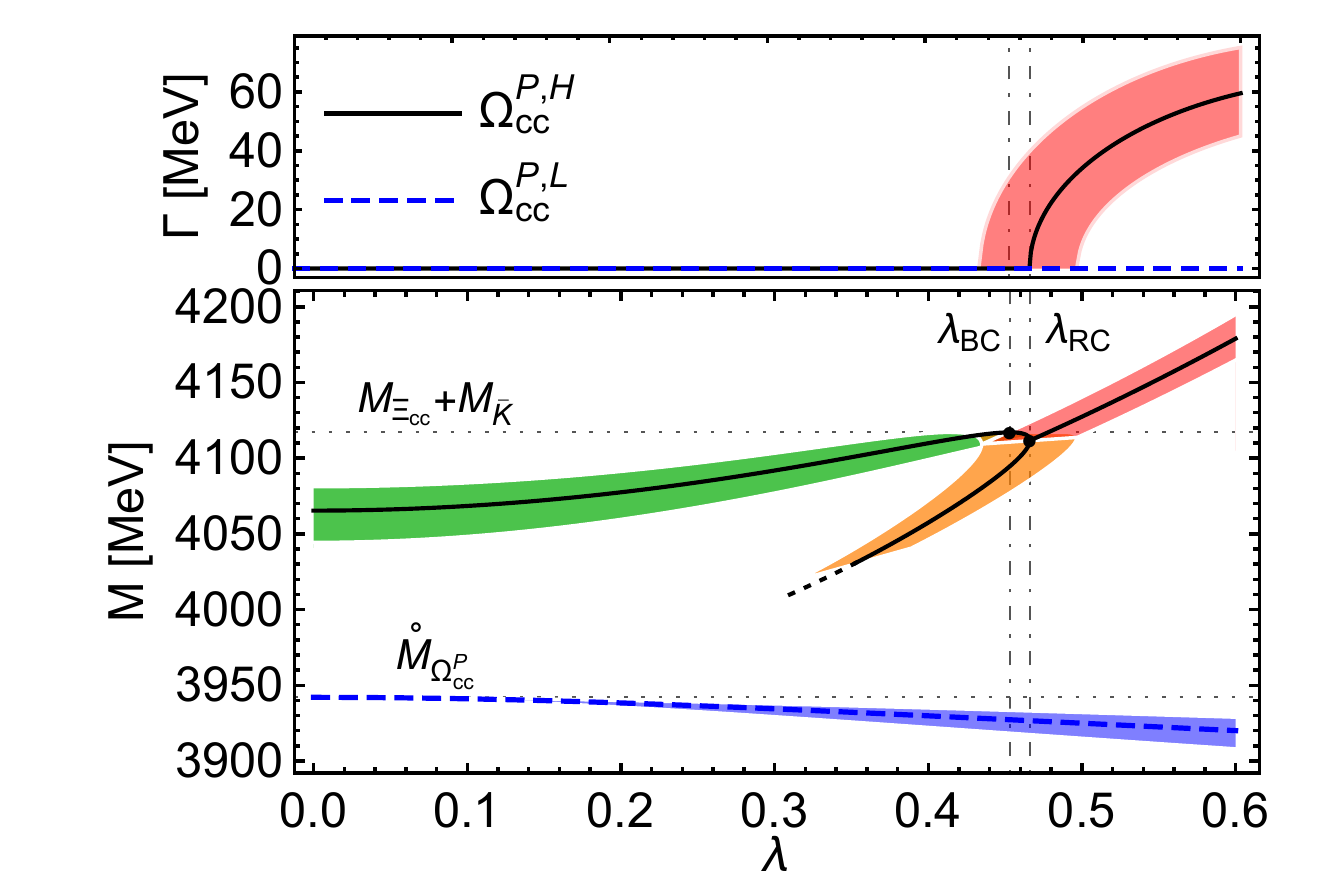}
	\caption{The widths (upper panel) and the masses of the two lowest $1/2^-$ $\Omega_{cc}^P$ states (lower panel) depending on the value of $\lambda$ with isospin symmetry imposed. The green, orange and red bands correspond to the cases of bound state, virtual states and resonance, respectively. 
The bands are obtained by taking into account uncertainties of the subtraction constant and the LECs determined in Ref.~\cite{Liu:2012zya}.} 
	\label{coupling-dependence}
\end{figure}
We first focus on the sector with $(S,I)=(-1,0)$ and $\lambda=0$. Then, in addition to the $\Omega_{cc}^{P}$ with a $P$-wave $cc$ excitation, one finds a pole below the $\Xi_{cc}\bar K$ threshold from the $\Xi_{cc}\bar K$--$\Omega_{cc}\eta$ coupled channel dynamics at about 4.07~GeV, analogous to the $D_{s0}^*(2317)$. The pole couples dominantly to $\Xi_{cc}\bar K$.
As long as $\lambda$ takes a nonvanishing value, as it should, the two states will mix with each other. It is expected that the state from the $P$-wave diquark excitation gets pushed down and the dynamically generated state is pushed up (denoted by $\omeL$ and $\omeH$, respectively). When $\lambda$ is larger than a critical value $\lambda_\text{BC}$, the higher pole $\omeH$ will change from a bound state to a virtual state. Increasing $\lambda$ further, $\omeH$ will become a resonance with the critical value denoted by $\lambda_\text{RC}$, see Fig.~\ref{coupling-dependence}. 
Such a behavior for an $S$-wave pole has already been observed in the study of the quark mass dependence of the lightest scalar meson $f_0(500)$~\cite{Hanhart:2008mx} and of the scalar charmed mesons~\cite{Guo:2009ct,Du:2017ttu}. The mass of $\omeL$ decreases monotonically. 
As already discussed, the natural value for $\lambda$ should be $\mathcal{O}(\Lambda_\text{QCD}/m_c) =  \mathcal{O}(0.2)$. From Fig.~\ref{coupling-dependence}, one sees that if $\lambda\lesssim 0.45$, both $1/2^-$ $\Omega_{cc}^P$ states are below the $\Xi_{cc}\bar K$ threshold. In this case, the only allowed strong decay mode is $\Omega_{cc}\pi^0$ which breaks isospin symmetry. Therefore, both states are expected to be very narrow. 

For an $S$-wave bound state with a small binding energy, the so-called compositeness~\cite{Weinberg:1962hj,Weinberg:1965zz,Baru:2003qq,Gamermann:2009uq,Hyodo:2013nka,Guo:2017jvc} 
measures the probability of finding the composite component in the wave function of the physical state. Here, one can evaluate the $\Xi_{cc}\bar{K}$ compositeness in $\omeH$ by using $-g_{\Xi_{cc}\bar{K}}^2 {\partial G_{\Xi_{cc}\bar{K}}}/{\partial s}$ at the pole of $\omeH$, where $g_{\Xi_{cc}\bar{K}}^2$ is the residue of the $T$-matrix element for the elastic $\Xi_{cc}\bar{K}$ channel. It is found that $\omeH$ contains around 55\%--80\% of $\Xi_{cc}\bar{K}$ when it is below the $\Xi_{cc}\bar{K}$ threshold.

If we use different values for the so far unobserved doubly charmed baryons, numerical results will change. However, the general mixing picture shown in Fig.~\ref{coupling-dependence} remains. 
For instance, the critical value $\lambda_\text{BC}$ changes to 0.40  if we increase $\mathring{M}_{\Omega_{cc}^P}$ by 40~MeV and keep all the other masses fixed. This is consistent with the expectation that the closer $\mathring{M}_{\Omega_{cc}^P}$ to the dynamically generated pole the stronger the mixing and thus the smaller $\lambda_\text{BC}$.

\begin{figure}[tb]
	\centering
	\includegraphics[width=0.50\textwidth]{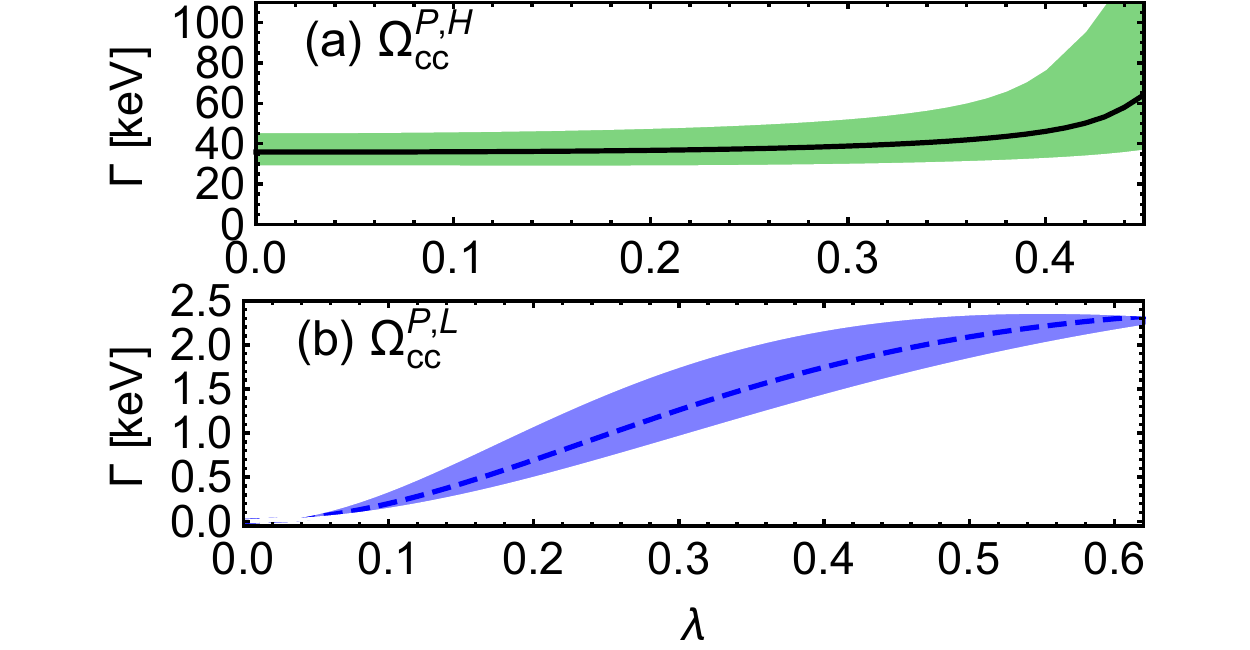} \\
	\caption{Isospin symmetry-breaking decay width of the higher $\omeH$ (a) and the lower $\omeL$ (b). } 
    \label{Isospin-breaking-HL}
\end{figure}

An anomalously large isospin-breaking partial decay width $\Gamma (D_{s0}^{*}(2317)\to D_{s}^+\pi^0 )$ of about 100~keV~\cite{Lutz:2007sk, Faessler:2007gv,Guo:2008gp,Liu:2012zya,Cleven:2014oka,Guo:2018kno} can be taken as an evidence for the $D_{s0}^{*}(2317)$ to be mainly a $D{K}$ molecule rather than a $P$-wave $c\bar s$ meson. This prediction will be checked at the $\overline{\text{P}}$ANDA experiment~\cite{Lutz:2009ff}.
Similarly, once the $1/2^-$ $\Omega_{cc}^{P}$ states will be discovered, one expects their isospin-breaking decay widths to be also important to reveal their nature.
The reason is that in the hadronic molecule case, the isospin mass splittings of the constituent hadrons play a dominant role in driving an isospin-breaking decay width much larger than the one generated by the $\pi^0$-$\eta$ mixing only. 
In order to calculate these tiny widths, one needs to work in the particle basis instead of the isospin basis. There are four channels: $\Omega_{cc}^{+}{\pi}^0,\ \Xi_{cc}^{++}K^{-},\ \Xi_{cc}^{+}\bar{K}^{0},$ and $\Omega_{cc}^{+}{\eta}$. 
We take the central values of all the meson masses from Ref.~\cite{Patrignani:2016xqp}, and $M_{\Xi_{cc}^{++}}-M_{\Xi_{cc}^{+}}=(2.16\pm 0.20)\ \mbox{MeV}$ from a lattice QCD computation~\cite{Borsanyi:2014jba}. Note that due to the interference between the electromagnetic and $m_d-m_u$ contributions~\cite{Brodsky:2011zs}, $M_{\Xi_{cc}^{++}}$ is a bit larger than  $M_{\Xi_{cc}^{+}}$.  This implies that the $\Xi_{cc}$ and kaon isospin splittings contribute in opposite directions, so that the isospin-breaking decay width of the $\omeH$ should be smaller than that of the $D_{s0}^{*}(2317)$ when $\lambda=0$. This expectation is confirmed by the explicit 
calculations as shown in Fig.~\ref{Isospin-breaking-HL}. It is found that the lower $\omeL$ gets a width of a few keV, while the width for the higher $\omeH$ is larger than $30$~keV.
The error bands in Fig.\ref{Isospin-breaking-HL} come from the uncertainties of the subtraction constant, of the LECs and of $M_{\Xi_{cc}^{++}}-M_{\Xi_{cc}^{+}}$.

\begin{figure}[tb]
	\centering
	\includegraphics[width=0.47\textwidth]{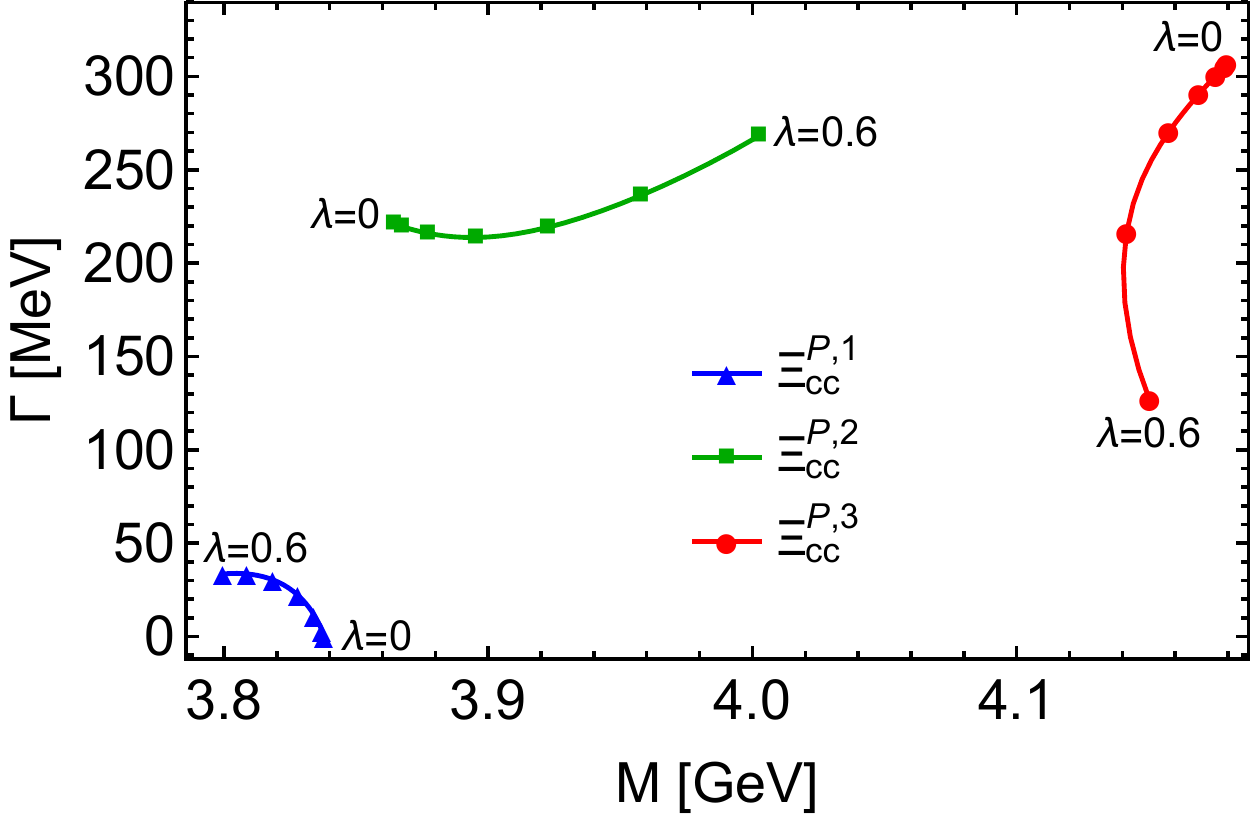}
	\caption{Trajectories of the three resonance poles in the $(S, I)=(0, 1/2)$ channel by changing the $\lambda$ value. Central values of LECs and $a^{\psi_{cc}\phi}$ are adopted, and $\mathring{M}_{\Xi_{cc}^P}=3838$~MeV~\cite{Ebert:2002ig} is used.}  \label{Isospin-one-half}
\end{figure}

Now let us turn to the sector with $(S,I)=(0,1/2)$ in the isospin symmetric limit.  Three resonance poles are found in the complex energy plane. Their positions with different $\lambda$ values are displayed in Fig.~\ref{Isospin-one-half}, where $\mathring{M}_{\Xi_{cc}^P}=3838$~MeV~\cite{Ebert:2002ig} is used. As can be seen, the lowest pole $\Xi_{cc}^{P,1}$ originates from the $P$-wave $cc$ excitation, and it has a small width less than 40~MeV. The seeds of the two broad poles $\Xi_{cc}^{P,2}$ and $\Xi_{cc}^{P,3}$ are the doubly charmed baryon counterparts of the two poles found in the coupled channel $D\pi$, $D\eta$ and $D_s\bar{K}$ scattering amplitudes~\cite{Hofmann:2003je,Guo:2006fu,Albaladejo:2016lbb,Du:2017zvv} belonging to the SU(3) flavor triplet and antisextet, respectively. 
Analogously, $\Xi_{cc}^{P,2}$ and $\Xi_{cc}^{P,3}$ couple most strongly to $\Xi_{cc}\pi$ and $\Omega_{cc}K$, respectively. Increasing $\lambda$ will make $M_{\Xi_{cc}^{P,1}}$ and $M_{\Xi_{cc}^{P,3}}$ smaller and push $M_{\Xi_{cc}^{P,2}}$ to larger values. When $\lambda$ is small, the masses of ${\Xi_{cc}^{P,1}}$ and ${\Xi_{cc}^{P,2}}$ are close. Therefore, in experiments where these particles can be produced, one would expect to see in the $\Xi_{cc}\pi$ invariant mass distribution a narrow peak on top of a broad bump. Depending on the interference from coupled channels, there might also be a dip. The only allowed strong decay channel for both ${\Xi_{cc}^{P,1}}$ and ${\Xi_{cc}^{P,2}}$ is $\Xi_{cc}\pi$. The natural channel to search for them is the $\Xi_{cc}^{++}\pi^-$. Presumably, the values of $\lambda$ and the
bare masses will be first determined from measuring the masses and widths of the lowest $\Xi_{cc}^{P}$ states. Then the rest of the spectrum can be predicted.

Note that in the results presented no corrections
to the assumed HADS were included. Those corrections can lead to 
variations of $\mathcal{O}(r_d/r_q)\sim 30\%$ in the LECs of the NLO interactions. While this in principle can lead to moderate
quantitative deviations from the predictions given above, these corrections
should not change the overall picture that is dominated by
the leading interactions, fixed completely by the chiral symmetry
of QCD, and the interplay with the $s$-channel poles.

In summary, 
we investigated the low-lying spectrum of the doubly charmed baryons with $J^P=1/2^-$ by studying the $S$-wave $\psi_{cc}\phi$ interactions in channels with $(S, I)=(-1, 0)$ and $(S, I)=(0, 1/2)$ using a unitarized coupled channel approach based on chiral effective Lagrangians up to NLO. The HADS is used to relate the NLO parameters to those in the charmed meson sector which have already been fixed and tested. The essential new point in this paper is that, in addition to the meson-baryon channels, the $P$-wave $cc$ diquark excitations have to be taken into account as dynamical degrees of freedom. As a result, the spectrum of $1/2^-$ doubly charmed baryons becomes richer than that known for positive-parity charmed and bottom mesons, and is also predicted to be different than predictions from quark models. 
The numerical results depend on inputs for the unobserved doubly baryon masses, of which rough estimates are known, and on one unknown coupling $\lambda=\mathcal{O}(\Lambda_\text{QCD}/m_c)\ll1$. When $\lambda\lesssim0.45$, which is likely, there exist two $1/2^-$ $\Omega_{cc}^P$ whose only strong decay mode is the isospin breaking $\Omega_{cc}\pi^0$.
Thus, both states should be very narrow. In the $(S,I)=(0,1/2)$ sector there are three $1/2^-$ $\Xi_{cc}^P$ states below 4.2~GeV. The lowest one has a narrow width while the other two are rather broad. We suggest to search for the lower states in the $\Xi_{cc}^{++}\pi^-$ decay mode. It is expected that the $3/2^-$ doubly charmed baryons and the $(1/2,3/2)^-$ doubly bottom and charm-bottom baryons possess the same pattern.

Searching for these particles and their analogues in future experiments will be helpful to establish the proper paradigm for excited hadrons. Given that LHCb already observed the $\Xi_{cc}^{++}$, we expect to see more exciting results in the near future on doubly charmed baryons.

\begin{acknowledgments}
Helpful discussions with Andreas Wirzba and De-Liang Yao are gratefully acknowledged. We thank Zhen-Wei Yang for discussions on the experimental perspective, and thank Pedro Fernandez-Soler for pointing out a typo in a previous version. This work is supported by the National Natural Science Foundation of China (NSFC) and Deutsche Forschungsgemeinschaft (DFG) through 
funds provided to the Sino--German Collaborative Research Center ``Symmetries and the Emergence of Structure in QCD'' (NSFC Grant No.~11621131001,
DFG Grant No.~TRR110), by the NSFC (Grant No.~11747601), by the Thousand Talents Plan for Young Professionals, by the CAS Key Research Program of Frontier Sciences
(Grant No.~QYZDB-SSW-SYS013), by the CAS Key Research Program (Grant
No. XDPB09), by the CAS President's
International Fellowship Initiative (PIFI) (Grant Nos.~2017PM0031
and 2018DM0034),  by the CAS Center for Excellence in Particle Physics (CCEPP), and by the VolkswagenStiftung (Grant No.~93562).
\end{acknowledgments}

\end{document}